\documentclass[sigconf,natbib=true]{acmart}

\usepackage{acronym}
\usepackage{amsmath}
\usepackage{multirow}
\usepackage{booktabs}
\usepackage[subrefformat=parens,labelformat=parens]{subfig}
\usepackage{xspace}

\newcommand{\msmarco}{\textsc{Ms Marco}\xspace}
\newcommand{\hqa}{\textsc{HotpotQA}\xspace}
\newcommand{\fever}{\textsc{Fever}\xspace}

\DeclareMathOperator*{\argmax}{arg\,max}

\AtBeginDocument{%
  \providecommand\BibTeX{{%
    \normalfont B\kern-0.5em{\scshape i\kern-0.25em b}\kern-0.8em\TeX}}}

\copyrightyear{2024}
\acmYear{2024}
\setcopyright{rightsretained}
\acmConference[SIGIR '24]{Proceedings of the 47th International ACM SIGIR Conference on Research and Development in Information Retrieval}{July 14--18, 2024}{Washington, DC, USA}
\acmBooktitle{Proceedings of the 47th International ACM SIGIR Conference on Research and Development in Information Retrieval (SIGIR '24), July 14--18, 2024, Washington, DC, USA}
\acmDOI{10.1145/3626772.3657931}
\acmISBN{979-8-4007-0431-4/24/07}

\begin{document}

\title{A Learning-to-Rank Formulation of Clustering-Based Approximate Nearest Neighbor Search}

\author{Thomas Vecchiato}
\affiliation{%
	\institution{Ca' Foscari University of Venice}
	\city{Venice}
	\country{Italy}
}
\email{880038@stud.unive.it}

\author{Claudio Lucchese}
\affiliation{%
	\institution{Ca' Foscari University of Venice}
	\city{Venice}
	\country{Italy}
}
\email{claudio.lucchese@unive.it}

\author{Franco Maria Nardini}
\affiliation{%
	\institution{ISTI-CNR}
	\city{Pisa}
	\country{Italy}
}
\email{francomaria.nardini@isti.cnr.it}

\author{Sebastian Bruch}
\affiliation{%
	\institution{Pinecone}
	\city{New York, NY}
	\country{USA}
}
\email{sbruch@acm.org}



\begin{abstract}
A critical piece of the modern information retrieval puzzle is
approximate nearest neighbor search. Its objective is to return a set of $k$
data points that are closest to a query point, with its accuracy measured by
the proportion of exact nearest neighbors captured in the returned set.
One popular approach to this question is clustering:
The indexing algorithm partitions data points into non-overlapping subsets and
represents each partition by a point such as its centroid.
The query processing algorithm first identifies the nearest clusters---a
process known as \emph{routing}---then performs a nearest neighbor search over
those clusters only.
In this work, we make a simple observation: The routing function
solves a ranking problem. Its quality can therefore be assessed with a ranking metric,
making the function amenable to learning-to-rank. Interestingly, ground-truth is often freely available:
Given a query distribution in a top-$k$ configuration, the ground-truth is the set of clusters
that contain the exact top-$k$ vectors.
We develop this insight and apply it to Maximum Inner Product Search (MIPS).
As we demonstrate empirically on various datasets, learning a simple linear function 
consistently improves the accuracy of clustering-based MIPS.
\end{abstract}

\ccsdesc[500]{Information systems~Retrieval models and ranking}
\keywords{Approximate Nearest Neighbor Search; Inverted File; Learning to Rank}

\maketitle


\section{Introduction}
\label{sec:introduction}
Information Retrieval (IR) is no stranger to the problem of top-$k$ retrieval.
Inverted indexes, for example, have long been the backbone of exact and approximate search
over \emph{sparse} vectors~\cite{tonellotto2018survey}, such as encodings of text using BM25~\cite{bm25original}.
With the advent of representation learning and the proliferation of dense vectors
as embeddings of text, a different form of approximate retrieval that originates
in neighboring scientific disciplines has become a major part of modern IR: Approximate
Nearest Neighbor (ANN) search~\cite{bruch2024foundations}.

The ANN setup should be familiar: A collection of vectors are processed
into a data structure, known as the index, in an offline phase.
When a query vector is presented, the ANN algorithm uses the index
to quickly find the closest vectors to the query point,
where closeness is defined based on the inner product between vectors or their Euclidean distance.
Importantly, the returned set is often an approximate solution: If $\mathcal{S}$ is the exact set of
$k$ closest vectors to the query, and $\tilde{\mathcal{S}}$ is the set returned by the ANN
algorithm, then the overlap between the two sets serves as a barometer quantifying the
accuracy of the search: $\lvert \mathcal{S} \cap \tilde{\mathcal{S}} \rvert / k$.

This work concerns ANN over dense vectors, with a particular focus on clustering-based methods,
where a collection is first split into $L$ geometric partitions using a clustering algorithm.
Each partition is then represented by a vector (e.g., its centroid).
The list of partitions along with their centroids make up the index.

At search time, the algorithm first computes the similarity---in this work,
inner product---between centroids and the query. It then identifies the $\ell$ closest
partitions and performs ANN search over those partitions only. We call the first step ``routing''
where a query vector is mapped to $\ell$ partitions. Using $\ell \ll L$ reduces the
search space dramatically, leading to an efficient search but at the expense of accuracy.
In this work, we assume that search within clusters is exact; in practice that itself
may be another ANN algorithm.

In that context, we observe a simple but consequential fact:
Given a fixed partitioning of the data, the routing machinery is a ranking function.
To see how, consider the top-$1$ case.
Given a query, the algorithm ranks partitions by their likelihood of containing the
nearest neighbor. As we assumed the clustering algorithm gives a \emph{partitioning}---so
that, a data vector belongs to one partition---a query will have
a single ``relevant'' answer: the partition that contains its nearest neighbor. Therefore,
routing quality can be determined with ranking metrics such as Mean Reciprocal Rank (MRR).

Perhaps more interesting is the fact that such a ranking function can be \emph{learnt} using
Learning-to-Rank (LTR)~\cite{bruch2023fntir}.
In fact, because MRR is an appropriate metric, and because each query has a single relevant answer,
the cross-entropy loss~\cite{bruch2019xendcg,bruch2021xendcg} is provably the right
surrogate to optimize. Furthermore, training data is cheap: All one requires is a set of queries. The ground-truth
``relevance'' labels can be computed by performing an exact search for each query, finding the nearest
neighbor, and identifying the partition it belongs to.

It appears that we have all the ingredients to learn a routing function for a fixed partitioning of vectors.
In this work, we develop that insight into a working solution.
We show through experiments on embeddings of various text datasets,
that learning a routing function can boost the accuracy of ANN search.
In fact, the gains are consistent regardless of our choice of the clustering algorithm.

What is even more appealing about our results is that we simply learn a linear
function: $\tau(q;\; W) = Wq$, where $q \in \mathbb{R}^d$ and $W \in \mathbb{R}^{L \times d}$.
The output of $\tau(\cdot)$ gives us the ranking scores for each of the $L$ partitions.
Pleasantly, that means that the $i$-th row of $W$ can be thought of as a learnt representative vector
of the $i$-th partition. That entails, we can conveniently plug in the learnt representatives into
the original routing machinery without any change whatsoever to implementation---an important criterion for production systems.

Our findings show the potential LTR holds for ANN search. We are excited to bridge these two fields
in our work and thereby motivate the community to explore this junction.
To that end, we briefly review the literature in Section~\ref{sec:related}, describe our method
in Section~\ref{sec:methodology}, and present experimental results in Section~\ref{sec:experiments}.
We conclude this work in Section~\ref{sec:conclusions} with remarks on possible future directions.


\section{Related Work}
\label{sec:related}

ANN algorithms come in many flavors including trees~\cite{kdtree,dasgupta2015rptrees}, hash tables~\cite{lsh},
and graphs~\cite{hnsw2020,diskann}. The method relevant to this work is the clustering-based
approach, also known as Inverted File (IVF)~\cite{pq}. As we intend to keep this review concise,
we refer the interested reader to~\cite{bruch2024foundations} for a thorough treatment of the subject.

As explained in Section~\ref{sec:introduction}, indexing in clustering-based ANN
utilizes a clustering algorithm to partition the vectors. A typical choice~\cite{pq,auvolat2015clustering}
is standard KMeans, which iterates between an expectation phase (computing partition centroids given
current cluster assignments) and an assignment step (computing cluster assignments given centroids).
This choice makes sense especially in ANN with Euclidean distance, as it can be shown that the KMeans
objective helps find Voronoi regions---a crucial structure in ANN search.

In this work, we also use two alternative clustering strategies. One is a variant of KMeans
known as Spherical KMeans~\cite{dhillon1999conceptde}, where after each expectation phase, the centroids
are $L_2$-normalized and projected onto the unit sphere. This small modification makes the clustering
algorithm arguably more suitable for ANN with angular distance and has been shown~\cite{bruch2023bridging}
to perform well for inner product.

The final clustering method used in this work is~\cite{chierichetti2007clusterPruning},
which we refer to as Shallow KMeans, and can be thought of as the first iteration of KMeans.
The algorithm begins by selecting $L$ data points uniformly at random and using them as cluster representatives.
It then assigns data points to partitions by inner product with the cluster representatives:
the partition whose representative yields the largest inner product with a data point becomes
its home partition.

We now briefly review LTR, a well-researched topic in IR.
In general, given a set of training queries with ground-truth (relevance) labels,
and a collection of items, the goal is to learn a function that computes a score
for any query-item pair. These scores induce an order among the items, resulting in a ranking.
The rankings generated by a learnt function ideally maximize a ranking metric
for a query distribution of interest, such as NDCG~\cite{jarvelin2002cumulated} or MRR.

Because ranking metrics are discontinuous or otherwise flat almost everywhere, supervised
LTR resorts to optimizing surrogate objectives~\cite{joachims2006training,Jun+Hang:2007,Rudin:JMLR:2009,RuWa:aistats:2018,Taylor+al:2008,qin2010general,burges2010ranknet,xia2008listwise,cao2007learning,burges2005learning,BruchApproxSIGIR2019}.
One objective that is appropriate for our work is the cross entropy-based loss
described in~\cite{bruch2021xendcg}. It has been shown~\cite{bruch2019xendcg,bruch2021xendcg}
that the cross-entropy loss is a consistent surrogate for MRR when each query has at most one
relevant item with probability $1$. Noting that our setup satisfies these conditions,
as explained in Section~\ref{sec:introduction}, we will use this ranking loss in this work.


\section{Methodology}
\label{sec:methodology}

\newcommand{\data}{\ensuremath{\mathcal{X}}}
\newcommand{\queries}{\ensuremath{\mathcal{Q}}}
\newcommand{\cluster}{\ensuremath{\mathcal{C}}}
\newcommand{\clustering}{\ensuremath{\Gamma}}
\newcommand{\universe}{\ensuremath{\mathbb{R}^d}}
\newcommand{\route}{\ensuremath{\tau}}

\acrodef{MIPS}[MIPS]{Maximum Inner Product Search}
\acrodef{MRR}[MRR]{Mean Reciprocal Rank}

Let $\left\langle u,v \right\rangle$ denote the inner product of two vectors $u,v \in \universe$. 
Given a collection of datapoints $\data \subset \universe$ and a query point $q \in \universe$, 
the (top-$1$) \ac{MIPS} problem is to find:
\begin{equation}
    \label{equation:mips}
    u^\ast = \argmax_{u \in \data} \left\langle q, u \right\rangle.
\end{equation}

As noted earlier, to solve this problem efficiently (albeit approximately), one prominent approach clusters \data{}
into a set of non-overlapping partitions $\{\cluster_i \}_{i=1}^L$, where $\cluster_i \in 2^\data$,
each represented with a vector $\mu_i$.
A routing function $\route:\universe \to \mathbb{R}^L$ then computes a ``goodness'' score for
each $C_i$. Typically, $\route=Mq$ where $M \in \mathbb{R}^{L \times d}$ whose $i$-th row is $\mu_i$.
Subsequently, only the points within the top $\ell$ clusters by score are evaluated
to find the solution to Equation~(\ref{equation:mips}).

We claim in this work that the function $\route(\cdot)$ can be learnt using supervised LTR methods.
Concretely, we wish to learn $\route(q;\; W)=Wq$ where $W \in \mathbb{R}^{L \times d}$.
We choose to learn a linear function
so that it can trivially replace $\route=Mq$ by setting $M=W$ without any change to the framework.
However, the method can be extended to any other parameterized family of functions.

To complete the picture, we must define a training dataset and a loss function.
We build training examples as follows. For query $q$, let $\route^\ast$ be the oracle routing function
such that the $i$-th component of $\route^\ast(q)$ is $1$ if $u^\ast \in C_i$ and $0$ otherwise.
The training set then comprises pairs $(q_i, \route^\ast(q_i))$, where $q_i \in \queries$ is a query point. 

We learn $\route(\cdot)$ by maximizing MRR. Note that MRR is especially meaningful, as opposed to
classification metrics, in the more general case where we allow probing $\ell>1$ partitions.
In this case, promoting a ranking function that places the relevant cluster among the top positions
increases the probability of finding $u^\ast$.

We use the cross-entropy loss to maximize MRR~\cite{bruch2019xendcg,bruch2021xendcg}.
The loss for a single query $q$ reduces to the following:
\begin{equation*}
    - \sum_{i=1}^L \route^\ast_i \log \frac{\exp(\route_i)}{ \sum_{j=1}^L \exp(\route_j)},
\end{equation*}
where, with a slight abuse of notation, we write $\route_k$ and $\route^\ast_k$ to denote
the $k$-th component of $\route(q)$ and $\route^\ast(q)$, respectively.
Note that, because $\route^\ast$ has a single non-zero component, the sum collapses to a single term.
The mean of the loss above over the entire training query set becomes the final objective for
the optimization problem.

\subsection{Generalizing to top-$k$}
\label{sec:methodology:top-k}
What we have described thus far concerns top-$1$ MIPS only. That is, for each query,
we have a single correct partition to route it to, and we wish to learn said function.
We use this setup in the rest of this work, but note that learning a routing
function that optimizes for top-$k$ with $k>1$ is straightforward. Let us explain why.

Suppose $k > 1$ and let $\mathcal{S}$ denote the (exact) set of top-$k$ data vectors
for query $q$. Then one example of an oracle routing function $\route^\ast(q)$ will have
$1$ in its $i$-th component if and only if $\mathcal{S} \cap C_i \neq \emptyset$,
and $0$ otherwise. The loss function, for query $q$, can then be updated to follow its general
from from~\cite{bruch2021xendcg} as follows:
\begin{equation*}
    - \sum_{i=1}^L \frac{2^{\route^\ast_i} - \gamma_i}{\sum_{j=1}^L 2^{\route^\ast_j} - \gamma_j}
    \log \frac{\exp(\route_i)}{ \sum_{j=1}^L \exp(\route_j)},
\end{equation*}
where, as before, $\route^\ast_k$ and $\route_k$ denote the $k$-th components of
$\route^\ast(q)$ and $\route(q)$, and $\gamma_k$'s are uniformly sampled from the unit interval.


\section{Experiments}
\label{sec:experiments}
We now report our experimental evaluation by
first detailing the methodology and then presenting and discussing the results.

\subsection{Experimental Setup}

\vspace{1mm}
\noindent \textbf{Datasets}.
We use three publicly-available datasets:
\msmarco~\cite{nguyen2016msmarco}, \hqa~\cite{yang2018hotpotqa}, and \fever~\cite{thorne2018fever}.
\msmarco Passage Retrieval consists of about $8.8$ million short passages
and $909{,}000$ train queries.
\hqa is a question answering dataset collected from the English Wikipedia,
consisting of $5.2$ million documents and $98{,}000$ queries.
Finally, \fever has $5.4$ million documents and $13{,}000$ queries.

\vspace{1mm}
\noindent \textbf{Embedding Models}.
We use embedding models to transform queries and documents into $d$-dimensional dense vectors.
The models include \texttt{tasB}\footnote{\url{https://huggingface.co/sentence-transformers/msmarco-distilbert-base-tas-b}}~\cite{tasb} 
($d$=768), and \texttt{contriever}\footnote{\url{https://huggingface.co/facebook/contriever}}~\cite{izacard2022unsupervised} ($d$=768).
We also include models fine-tuned by~\cite{reimers-gurevych-2019-sentence}
on $1$ billion text pairs:
\texttt{all-MiniLM-L6-v2}\footnote{\url{https://huggingface.co/sentence-transformers/all-MiniLM-L6-v2}} ($d$=384),
\texttt{all-mpnet-base-v2}\footnote{\url{https://huggingface.co/sentence-transformers/all-mpnet-base-v2}} ($d$=768),
and \texttt{all-distilroberta-v1}\footnote{\url{https://huggingface.co/sentence-transformers/all-distilroberta-v1}} ($d$=768).

\vspace{1mm}
\noindent \textbf{Evaluation Metric}.
We evaluate all methods in terms of top-$k$ \emph{accuracy}, defined as follows.
For each query, we obtain the top $\ell$ partitions according to $\route(\cdot)$,
and record the percentage of the top-$k$ documents contained in those partitions.

\vspace{1mm}
\noindent \textbf{Baseline}.
We evaluate the accuracy of a learnt $\tau(\cdot)$ against the baseline routing function
$\tau(q)=Mq$ where the $i$-th row of $M$ is simply the representative of the $i$-th partition, 
$\mu_i$, returned by the clustering algorithm.
Specifically, we identify the top $\ell$ partitions for $q$ according to $Mq$,
and measure its accuracy as defined above.

\vspace{1mm}
\noindent \textbf{Implementation Details}.
We model $\route(\cdot;\; W)$ as a linear function with $W \in \mathbb{R}^{L \times d}$.
To compute the loss function, we transform $Wq$ with softmax,
which gives the probability of a query being routed to each partition.
We use Adam~\cite{kingma2017adam}
with a learning rate of $10^{-4}$ to optimize the loss function. 
We set the batch size to $512$ and train for a maximum of $100$
epochs.\footnote{Our code is available at: 
\url{https://github.com/tomvek/mips-learnt-ivf}}

Given a collection of $m$ vectors (i.e., embeddings of a dataset), 
we take the following steps in each experiment.
We first run one of the three clustering algorithms 
(standard, spherical, and shallow KMeans) on the data vectors,
clustering them into $L = \sqrt{m}$ partitions.
We then split the query set into training ($60\%$), validation ($20\%$), and
test ($20\%$) queries, and construct training, validation, 
and test examples using the procedure described in Section~\ref{sec:methodology}.
Finally, we train on the training examples 
and take the best model according to the loss on the validation examples.
We evaluate baseline and learnt functions on the test splits.

\begin{table}
\begin{center}
\begin{sc}
\footnotesize
\caption{\small Top-$1$ accuracy
of baseline and learnt routing functions on \texttt{all-MiniLM-L6-v2}
embeddings. Columns are blocked by the clustering algorithm.
For each dataset-clustering pair, we measure top-$1$ accuracy by setting $\ell$
to $0.1\%$ and $1\%$ of the number of partitions, $L$.
\label{tab:results:allMiniLM}}
\begin{tabular}{ll|cc|cc|cc}
\toprule
\multirow{2}{*}{Dataset} & \multirow{2}{*}{Method} & \multicolumn{2}{c}{Standard} & \multicolumn{2}{c}{Spherical} & \multicolumn{2}{c}{Shallow}\\

& & 0.1\% & 1\% & 0.1\% & 1\% & 0.1\% & 1\%\\

\midrule

\multirow{2}{*}{\msmarco}
& {Baseline} & 0.392 & 0.779 & 0.627 & 0.869 & 0.517 & 0.815\\
& {Learnt}   & 0.746 & 0.940 & 0.751 & 0.938 & 0.670 & 0.923\\
\midrule

\multirow{2}{*}{\hqa}
& {Baseline} & 0.089 & 0.481 & 0.328 & 0.684 & 0.258 & 0.724\\
& {Learnt}   & 0.488 & 0.844 & 0.493 & 0.833 & 0.412 & 0.827\\
\midrule

\multirow{2}{*}{\fever}
& {Baseline} & 0.102 & 0.443 & 0.249 & 0.562 & 0.279 & 0.621\\
& {Learnt}   & 0.663 & 0.865 & 0.662 & 0.872 & 0.633 & 0.912\\
\bottomrule

\end{tabular}
\end{sc}
\end{center}
\end{table}

\subsection{Experimental Results}

\noindent \textbf{Top-$1$}.
Table~\ref{tab:results:allMiniLM} presents the top-$1$ accuracy 
for baseline and learnt routing functions 
on the \texttt{all-MiniLM-L6-v2} embeddings of text datasets. 
As the table shows, we consider three configurations 
where the partitions are produced by standard, spherical and shallow KMeans.
For each configuration, we measure accuracy  by setting $\ell$ to $0.1\%$ 
and $1\%$ of $L$, the total number of partitions.

As is evident from Table~\ref{tab:results:allMiniLM},
a learnt $\tau(\cdot)$ consistently outperforms the baseline.
Taking standard KMeans and $\ell=L/100$ as an example, 
learning improves accuracy by about $21\%$ on \msmarco, 
$75\%$ on \hqa, and $95\%$ on \fever.
The difference is greater for smaller values of $\ell$,
indicating that the closest partitions to query vectors are of much greater quality.
That makes sense as our learning objective heavily focuses on the top partition.

We make another observation that may be of independent interest:
The shallow KMeans clustering algorithm performs remarkably well for MIPS.
Considering the fact that shallow KMeans requires no training
whatsoever---clustering is a matter of sampling $L$ points from the dataset 
followed by assignment---coupled with its high accuracy,
it makes for an efficient yet effective clustering in practice.

\begin{figure*}[t]
\begin{center}
\centerline{
\subfloat[\msmarco]{
    \includegraphics[width=0.33\textwidth]{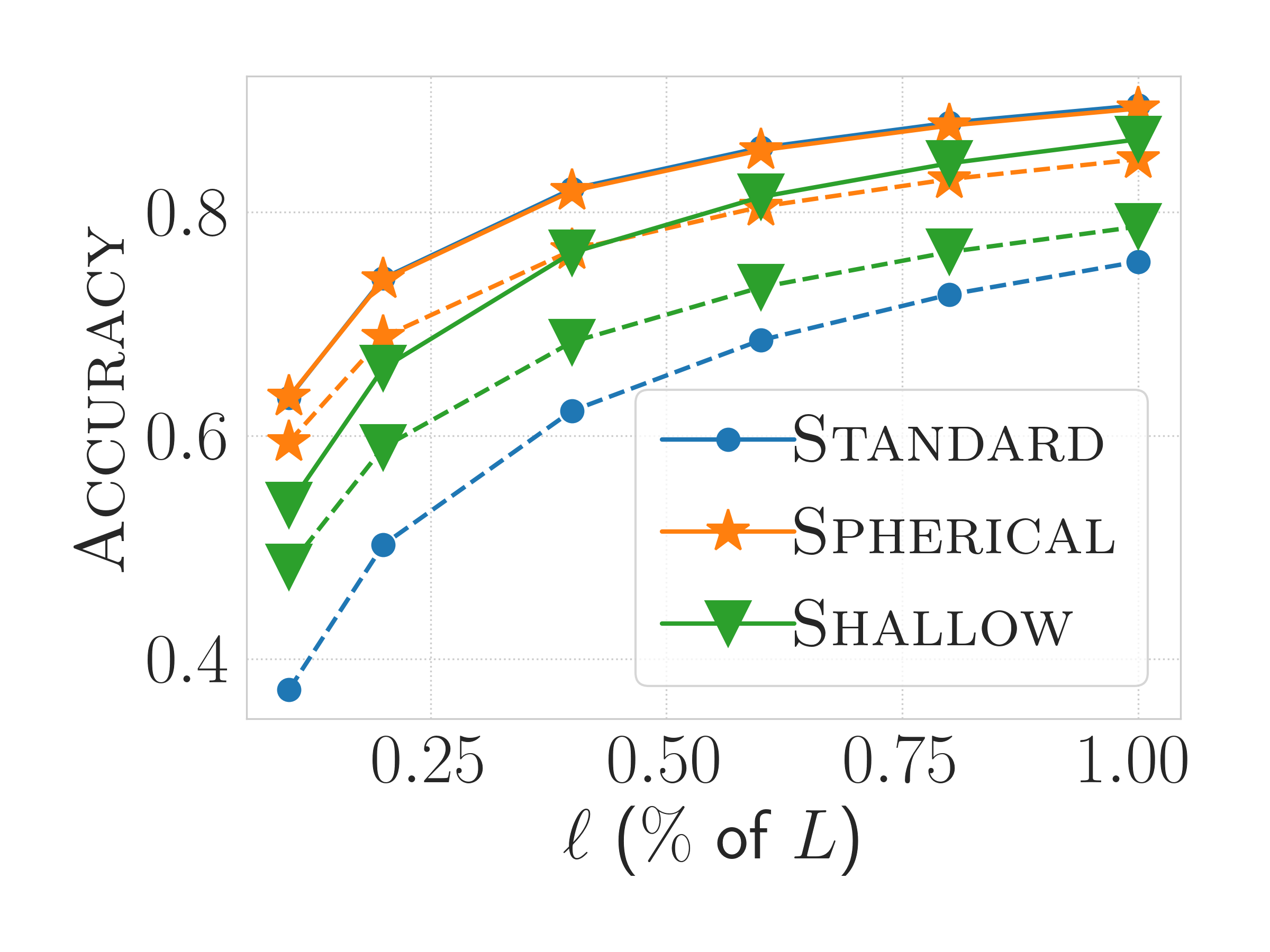}
}
\subfloat[\hqa]{
    \includegraphics[width=0.33\textwidth]{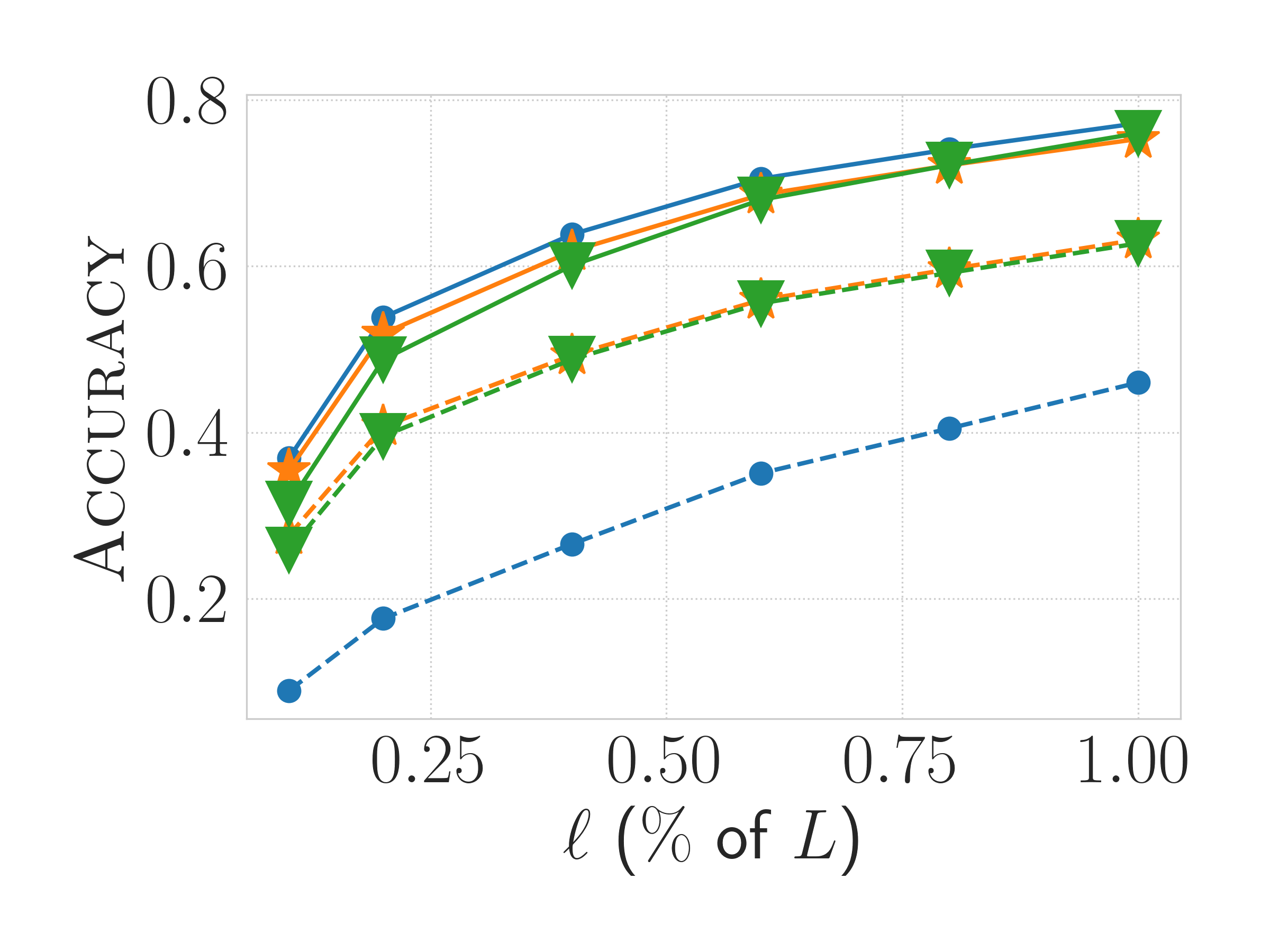}
}
\subfloat[\fever]{
    \includegraphics[width=0.33\textwidth]{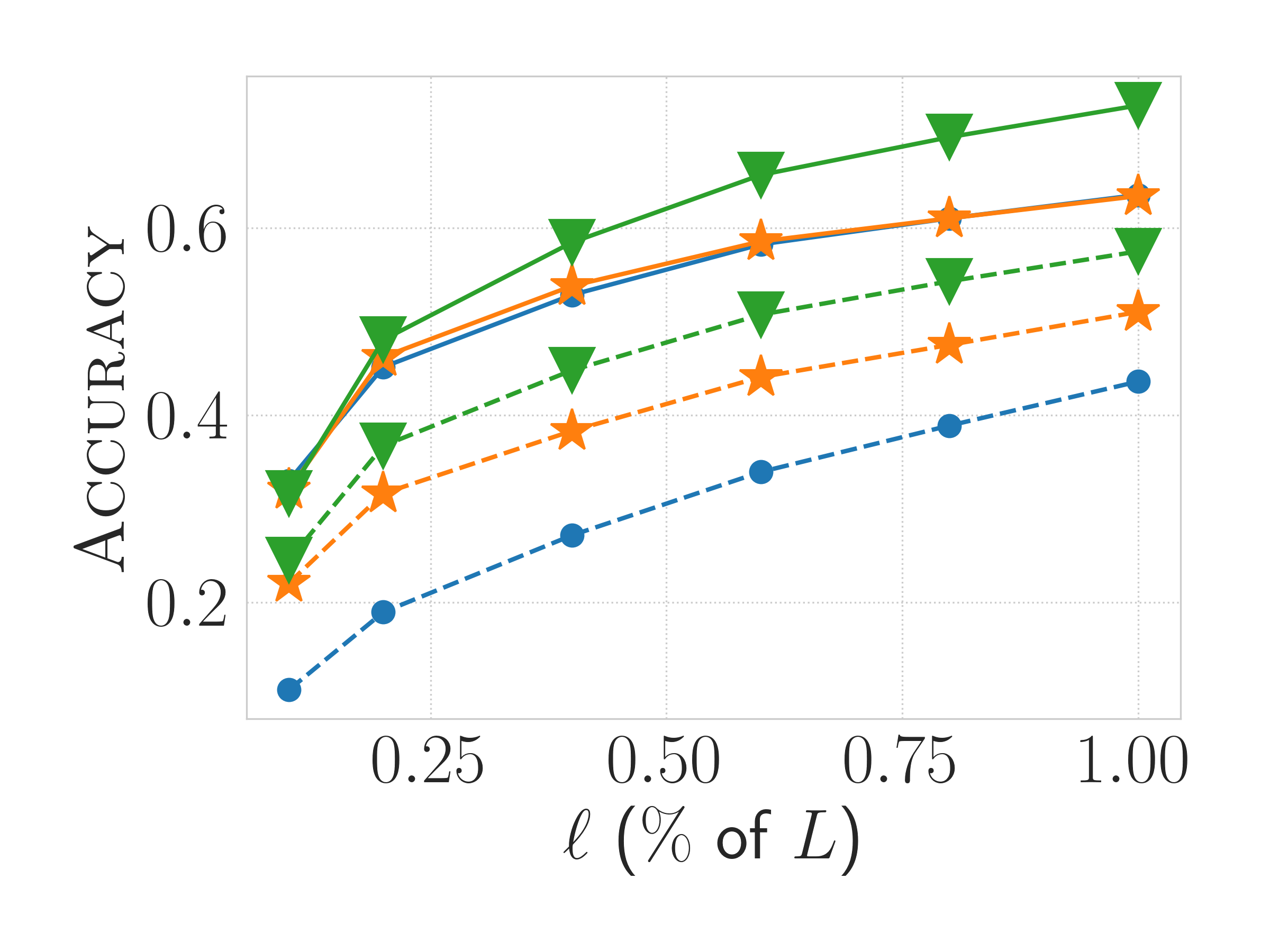}
}}

\caption{\small
Top-$10$ accuracy as a function of $\ell$ (expressed as percent of total number of partitions, $L$),
on the \texttt{all-MiniLM-L6-v2} embeddings.
In all figures, the dashed lines indicate the baseline and the solid lines show the performance of
the learnt routing function.
}
\label{figure:top-k}
\end{center}
\end{figure*}

\vspace{1mm}
\noindent \textbf{Top-$k$}.
Even though, when learning a routing function, 
we optimize a loss that is only concerned with top-$1$ accuracy,
we investigate in this section the impact of such training on top-$k$ accuracy.
Does learning to optimize top-$1$ accuracy benefit top-$k$ retrieval for $k>1$?
The answer appears to be in the affirmative, as the results of our experiments
show in Figure~\ref{figure:top-k},
rendering top-$10$ accuracy as a function of $\ell$. Trends are similar for $k = 100$.

These results are encouraging and bode well for an
extension of the loss function to top-$k$ retrieval, as explained in Section~\ref{sec:methodology:top-k}.
We leave an exploration of that generalization to future work.

\vspace{1mm}
\noindent \textbf{Other Embeddings}.
We repeat this exercise for other embedding models.
Table~\ref{tab:results:other-embeddings} reports the top-$1$ accuracy for select values
of $\ell$ on \msmarco. We observe that the general trend
from before holds: Learning does lead to gains in accuracy,
especially when $\ell$ is small.
The McNemar's test~\cite{mcnemartest1947} suggests statistically
significant difference between the methods ($p$-value $< 0.001$).

Unlike the previous results, gaps between the baseline and learnt routing
are smaller.
We believe what contributes to smaller gains
has to do with the larger dimensionality of the embedding vectors 
($768$ of models in Table~\ref{tab:results:other-embeddings}
versus $384$ of \texttt{all-MiniLM-L6-v2})
and the capacity of the linear function in learning appropriate partition representatives.
This is a question we leave to future work.

\begin{table}
\begin{center}
\begin{sc}
\footnotesize
\caption{
\small Top-$1$ accuracy of baseline and learnt routing on \msmarco.
For each embedding model, the top row represents the baseline accuracy 
and the bottom row the learnt function's.
As before, we report top-$1$ accuracy by setting $\ell$ to $0.1$ 
and $1$ percent of $L$.
\label{tab:results:other-embeddings}}
\begin{tabular}{l|cc|cc|cc}
\toprule
\multirow{2}{*}{Encoding} & \multicolumn{2}{c}{Standard} & \multicolumn{2}{c}{Spherical} & \multicolumn{2}{c}{Shallow}\\

& 0.1\% & 1\% & 0.1\% & 1\% & 0.1\% & 1\%\\

\midrule

\multirow{2}{*}{\texttt{tasB}}
& 0.480 & 0.835 & 0.680 & 0.915 & 0.553 & 0.869\\
& 0.727 & 0.936 & 0.724 & 0.933 & 0.612 & 0.896\\
\midrule

\multirow{2}{*}{\texttt{contriever}} 
& 0.602 & 0.895 & 0.756 & 0.938 & 0.640 & 0.909\\
& 0.790 & 0.952 & 0.780 & 0.946 & 0.690 & 0.927\\
\midrule

\multirow{2}{*}{\texttt{all-mpnet-base-v2}} 
& 0.763 & 0.952 & 0.794 & 0.958 & 0.691 & 0.940\\
& 0.818 & 0.967 & 0.819 & 0.966 & 0.733 & 0.951\\
\midrule

\multirow{2}{*}{\texttt{all-distilroberta-v1}} 
& 0.752 & 0.955 & 0.779 & 0.960 & 0.664 & 0.935\\
& 0.807 & 0.967 & 0.806 & 0.966 & 0.706 & 0.945\\
\bottomrule

\end{tabular}
\end{sc}
\end{center}
\end{table}


\section{Conclusions and Future Work}
\label{sec:conclusions}

Formalizing routing in a clustering-based ANN algorithm as ranking
is intuitive enough. When a partitioning has been finalized, the task is to rank
partitions by their likelihood of containing the nearest neighbor to a query.
Building on that intuition, we explored the feasibility of employing LTR algorithms
to improve routing. As we argued in this work, preparing the ingredients necessary
to apply LTR to routing, such as labeled training data, requires a modest
effort---in fact, all that is needed is a set of queries.

We empirically assessed the advantages of learning a (linear) routing function
and demonstrated the potential LTR holds for clustering-based ANN.
We observed that, generally, learning leads to gains in accuracy. In other words,
for the same number of partitions probed, a learnt function routes the query to
better partitions a higher percentage of the time.

We have only scratched the surface in this exploratory work,
with many interesting questions left to investigate in follow-up studies.
As hinted earlier, we wish to understand the impact of the more general loss
function of Section~\ref{sec:methodology:top-k} on top-$k$ retrieval.
Additionally, we hope to shed more light on the smaller gains on high-dimensional embeddings.
Finally, we intend to extend the method to other metrics such as Euclidean
or angular distance.

More exciting still is the question of the utility of LTR in the partitioning algorithm itself.
Recall that, throughout this work, we assumed that a collection of vectors has already been partitioned
by some clustering algorithm. We then ``refined'' the representative vectors through supervised learning.
What is clear, however, is that such learning can be incorporated into the clustering phase itself.
In other words, during clustering, one may iteratively refine the cluster representatives and
update cluster assignments accordingly. This opens the door to query-aware clustering for ANN
search---an area that has remained under-explored.

\vspace{1mm}
\noindent \textbf{Acknowledgements}.
This work was partially supported by the Horizon Europe RIA ``Extreme Food Risk Analytics'' (EFRA), grant agreement n. 101093026, by the iNEST - Interconnected Nord-Est Innovation Ecosystem (iNEST ECS\_00000043 – CUP H43C22000540006) and the PNRR - M4C2 - Investimento 1.3, Partenariato Esteso PE00000013 - ``FAIR - Future Artificial Intelligence Research'' - Spoke 1 ``Human-centered AI'' both funded by the European Commission under the NextGeneration EU program. The views and opinions expressed are solely those of the authors and do not necessarily reflect those of the European Union, nor can the European Union be held responsible for them.

\newpage

\bibliographystyle{ACM-Reference-Format}
\bibliography{biblio}


\begin{thebibliography}{36}


\ifx \showCODEN    \undefined \def \showCODEN     #1{\unskip}     \fi
\ifx \showDOI      \undefined \def \showDOI       #1{#1}\fi
\ifx \showISBNx    \undefined \def \showISBNx     #1{\unskip}     \fi
\ifx \showISBNxiii \undefined \def \showISBNxiii  #1{\unskip}     \fi
\ifx \showISSN     \undefined \def \showISSN      #1{\unskip}     \fi
\ifx \showLCCN     \undefined \def \showLCCN      #1{\unskip}     \fi
\ifx \shownote     \undefined \def \shownote      #1{#1}          \fi
\ifx \showarticletitle \undefined \def \showarticletitle #1{#1}   \fi
\ifx \showURL      \undefined \def \showURL       {\relax}        \fi
\providecommand\bibfield[2]{#2}
\providecommand\bibinfo[2]{#2}
\providecommand\natexlab[1]{#1}
\providecommand\showeprint[2][]{arXiv:#2}

\bibitem[Auvolat et~al\mbox{.}(2015)]%
        {auvolat2015clustering}
\bibfield{author}{\bibinfo{person}{Alex Auvolat}, \bibinfo{person}{Sarath
  Chandar}, \bibinfo{person}{Pascal Vincent}, \bibinfo{person}{Hugo
  Larochelle}, {and} \bibinfo{person}{Yoshua Bengio}.}
  \bibinfo{year}{2015}\natexlab{}.
\newblock \bibinfo{title}{Clustering is Efficient for Approximate Maximum Inner
  Product Search}.
\newblock
\newblock
\showeprint[arxiv]{1507.05910}~[cs.LG]


\bibitem[Bentley(1975)]%
        {kdtree}
\bibfield{author}{\bibinfo{person}{Jon~Louis Bentley}.}
  \bibinfo{year}{1975}\natexlab{}.
\newblock \showarticletitle{Multidimensional Binary Search Trees Used for
  Associative Searching}.
\newblock \bibinfo{journal}{\emph{Commun. ACM}} \bibinfo{volume}{18},
  \bibinfo{number}{9} (\bibinfo{date}{9} \bibinfo{year}{1975}),
  \bibinfo{pages}{509–517}.
\newblock


\bibitem[Bruch(2021)]%
        {bruch2021xendcg}
\bibfield{author}{\bibinfo{person}{Sebastian Bruch}.}
  \bibinfo{year}{2021}\natexlab{}.
\newblock \showarticletitle{An Alternative Cross Entropy Loss for
  Learning-to-Rank}. In \bibinfo{booktitle}{\emph{Proceedings of the Web
  Conference 2021}} (Ljubljana, Slovenia). \bibinfo{pages}{118–126}.
\newblock


\bibitem[Bruch(2024)]%
        {bruch2024foundations}
\bibfield{author}{\bibinfo{person}{Sebastian Bruch}.}
  \bibinfo{year}{2024}\natexlab{}.
\newblock \bibinfo{booktitle}{\emph{Foundations of Vector Retrieval}}.
\newblock \bibinfo{publisher}{Springer Nature Switzerland}.
\newblock
\showISBNx{9783031551826}


\bibitem[Bruch et~al\mbox{.}(2023a)]%
        {bruch2023fntir}
\bibfield{author}{\bibinfo{person}{Sebastian Bruch}, \bibinfo{person}{Claudio
  Lucchese}, {and} \bibinfo{person}{Franco~Maria Nardini}.}
  \bibinfo{year}{2023}\natexlab{a}.
\newblock \showarticletitle{Efficient and Effective Tree-based and Neural
  Learning to Rank}.
\newblock \bibinfo{journal}{\emph{Foundations and Trends in Information
  Retrieval}} \bibinfo{volume}{17}, \bibinfo{number}{1} (\bibinfo{year}{2023}),
  \bibinfo{pages}{1--123}.
\newblock


\bibitem[Bruch et~al\mbox{.}(2023b)]%
        {bruch2023bridging}
\bibfield{author}{\bibinfo{person}{Sebastian Bruch},
  \bibinfo{person}{Franco~Maria Nardini}, \bibinfo{person}{Amir Ingber}, {and}
  \bibinfo{person}{Edo Liberty}.} \bibinfo{year}{2023}\natexlab{b}.
\newblock \bibinfo{title}{Bridging Dense and Sparse Maximum Inner Product
  Search}.
\newblock
\newblock
\showeprint[arxiv]{2309.09013}~[cs.IR]


\bibitem[Bruch et~al\mbox{.}(2019a)]%
        {bruch2019xendcg}
\bibfield{author}{\bibinfo{person}{Sebastian Bruch}, \bibinfo{person}{Xuanhui
  Wang}, \bibinfo{person}{Michael Bendersky}, {and} \bibinfo{person}{Marc
  Najork}.} \bibinfo{year}{2019}\natexlab{a}.
\newblock \showarticletitle{An Analysis of the Softmax Cross Entropy Loss for
  Learning-to-Rank with Binary Relevance}. In
  \bibinfo{booktitle}{\emph{Proceedings of the 2019 ACM SIGIR International
  Conference on Theory of Information Retrieval}} (Santa Clara, CA, USA).
  \bibinfo{pages}{75–78}.
\newblock


\bibitem[Bruch et~al\mbox{.}(2019b)]%
        {BruchApproxSIGIR2019}
\bibfield{author}{\bibinfo{person}{Sebastian Bruch}, \bibinfo{person}{Masrour
  Zoghi}, \bibinfo{person}{Mike Bendersky}, {and} \bibinfo{person}{Marc
  Najork}.} \bibinfo{year}{2019}\natexlab{b}.
\newblock \showarticletitle{Revisiting Approximate Metric Optimization in the
  Age of Deep Neural Networks}. In \bibinfo{booktitle}{\emph{Proceedings of the
  42nd International ACM SIGIR Conference on Research and Development in
  Information Retrieval}}.
\newblock


\bibitem[Burges et~al\mbox{.}(2005)]%
        {burges2005learning}
\bibfield{author}{\bibinfo{person}{Chris Burges}, \bibinfo{person}{Tal Shaked},
  \bibinfo{person}{Erin Renshaw}, \bibinfo{person}{Ari Lazier},
  \bibinfo{person}{Matt Deeds}, \bibinfo{person}{Nicole Hamilton}, {and}
  \bibinfo{person}{Greg Hullender}.} \bibinfo{year}{2005}\natexlab{}.
\newblock \showarticletitle{Learning to rank using gradient descent}. In
  \bibinfo{booktitle}{\emph{Proceedings of the 22nd International Conference on
  Machine Learning}}. \bibinfo{pages}{89--96}.
\newblock


\bibitem[Burges(2010)]%
        {burges2010ranknet}
\bibfield{author}{\bibinfo{person}{Christopher~J.C. Burges}.}
  \bibinfo{year}{2010}\natexlab{}.
\newblock \bibinfo{booktitle}{\emph{From {RankNet} to {LambdaRank} to
  {LambdaMART}: An Overview}}.
\newblock \bibinfo{type}{{T}echnical {R}eport} MSR-TR-2010-82.
  \bibinfo{institution}{Microsoft Research}.
\newblock


\bibitem[Cao et~al\mbox{.}(2007)]%
        {cao2007learning}
\bibfield{author}{\bibinfo{person}{Zhe Cao}, \bibinfo{person}{Tao Qin},
  \bibinfo{person}{Tie-Yan Liu}, \bibinfo{person}{Ming-Feng Tsai}, {and}
  \bibinfo{person}{Hang Li}.} \bibinfo{year}{2007}\natexlab{}.
\newblock \showarticletitle{Learning to rank: from pairwise approach to
  listwise approach}. In \bibinfo{booktitle}{\emph{Proceedings of the 24th
  International Conference on Machine Learning}}. \bibinfo{pages}{129--136}.
\newblock


\bibitem[Chierichetti et~al\mbox{.}(2007)]%
        {chierichetti2007clusterPruning}
\bibfield{author}{\bibinfo{person}{Flavio Chierichetti},
  \bibinfo{person}{Alessandro Panconesi}, \bibinfo{person}{Prabhakar Raghavan},
  \bibinfo{person}{Mauro Sozio}, \bibinfo{person}{Alessandro Tiberi}, {and}
  \bibinfo{person}{Eli Upfal}.} \bibinfo{year}{2007}\natexlab{}.
\newblock \showarticletitle{Finding near Neighbors through Cluster Pruning}. In
  \bibinfo{booktitle}{\emph{Proceedings of the 26th ACM SIGMOD Symposium on
  Principles of Database Systems}} (Beijing, China). \bibinfo{pages}{103--112}.
\newblock


\bibitem[Dasgupta and Sinha(2015)]%
        {dasgupta2015rptrees}
\bibfield{author}{\bibinfo{person}{Sanjoy Dasgupta} {and}
  \bibinfo{person}{Kaushik Sinha}.} \bibinfo{year}{2015}\natexlab{}.
\newblock \showarticletitle{Randomized Partition Trees for Nearest Neighbor
  Search}.
\newblock \bibinfo{journal}{\emph{Algorithmica}} \bibinfo{volume}{72},
  \bibinfo{number}{1} (\bibinfo{date}{5} \bibinfo{year}{2015}),
  \bibinfo{pages}{237–263}.
\newblock


\bibitem[Dhillon and Modha(1999)]%
        {dhillon1999conceptde}
\bibfield{author}{\bibinfo{person}{Inderjit~S. Dhillon} {and}
  \bibinfo{person}{Dharmendra~S. Modha}.} \bibinfo{year}{1999}\natexlab{}.
\newblock \bibinfo{booktitle}{\emph{Concept Decompositions for Large Sparse
  Text Data using Clustering}}.
\newblock \bibinfo{type}{{T}echnical {R}eport} RJ 10147.
  \bibinfo{institution}{Array}.
\newblock


\bibitem[Hofst\"{a}tter et~al\mbox{.}(2021)]%
        {tasb}
\bibfield{author}{\bibinfo{person}{Sebastian Hofst\"{a}tter},
  \bibinfo{person}{Sheng-Chieh Lin}, \bibinfo{person}{Jheng-Hong Yang},
  \bibinfo{person}{Jimmy Lin}, {and} \bibinfo{person}{Allan Hanbury}.}
  \bibinfo{year}{2021}\natexlab{}.
\newblock \showarticletitle{Efficiently Teaching an Effective Dense Retriever
  with Balanced Topic Aware Sampling}. In \bibinfo{booktitle}{\emph{Proceedings
  of the 44th International ACM SIGIR Conference on Research and Development in
  Information Retrieval}} (Virtual Event, Canada). \bibinfo{pages}{113--122}.
\newblock


\bibitem[Indyk and Motwani(1998)]%
        {lsh}
\bibfield{author}{\bibinfo{person}{Piotr Indyk} {and} \bibinfo{person}{Rajeev
  Motwani}.} \bibinfo{year}{1998}\natexlab{}.
\newblock \showarticletitle{Approximate Nearest Neighbors: Towards Removing the
  Curse of Dimensionality}. In \bibinfo{booktitle}{\emph{Proceedings of the
  30th Annual ACM Symposium on Theory of Computing}} (Dallas, Texas, USA).
  \bibinfo{pages}{604--613}.
\newblock


\bibitem[Izacard et~al\mbox{.}(2022)]%
        {izacard2022unsupervised}
\bibfield{author}{\bibinfo{person}{Gautier Izacard}, \bibinfo{person}{Mathilde
  Caron}, \bibinfo{person}{Lucas Hosseini}, \bibinfo{person}{Sebastian Riedel},
  \bibinfo{person}{Piotr Bojanowski}, \bibinfo{person}{Armand Joulin}, {and}
  \bibinfo{person}{Edouard Grave}.} \bibinfo{year}{2022}\natexlab{}.
\newblock \showarticletitle{Unsupervised Dense Information Retrieval with
  Contrastive Learning}.
\newblock \bibinfo{journal}{\emph{Transactions on Machine Learning Research}}
  (\bibinfo{year}{2022}).
\newblock


\bibitem[J{\"a}rvelin and Kek{\"a}l{\"a}inen(2002)]%
        {jarvelin2002cumulated}
\bibfield{author}{\bibinfo{person}{Kalervo J{\"a}rvelin} {and}
  \bibinfo{person}{Jaana Kek{\"a}l{\"a}inen}.} \bibinfo{year}{2002}\natexlab{}.
\newblock \showarticletitle{Cumulated gain-based evaluation of IR techniques}.
\newblock \bibinfo{journal}{\emph{ACM Transactions on Information Systems}}
  \bibinfo{volume}{20}, \bibinfo{number}{4} (\bibinfo{year}{2002}),
  \bibinfo{pages}{422--446}.
\newblock


\bibitem[Jayaram~Subramanya et~al\mbox{.}(2019)]%
        {diskann}
\bibfield{author}{\bibinfo{person}{Suhas Jayaram~Subramanya},
  \bibinfo{person}{Fnu Devvrit}, \bibinfo{person}{Harsha~Vardhan Simhadri},
  \bibinfo{person}{Ravishankar Krishnawamy}, {and} \bibinfo{person}{Rohan
  Kadekodi}.} \bibinfo{year}{2019}\natexlab{}.
\newblock \showarticletitle{DiskANN: Fast Accurate Billion-point Nearest
  Neighbor Search on a Single Node}. In \bibinfo{booktitle}{\emph{Advances in
  Neural Information Processing Systems}}, Vol.~\bibinfo{volume}{32}.
\newblock


\bibitem[J\'egou et~al\mbox{.}(2011)]%
        {pq}
\bibfield{author}{\bibinfo{person}{Herve J\'egou}, \bibinfo{person}{Matthijs
  Douze}, {and} \bibinfo{person}{Cordelia Schmid}.}
  \bibinfo{year}{2011}\natexlab{}.
\newblock \showarticletitle{Product Quantization for Nearest Neighbor Search}.
\newblock \bibinfo{journal}{\emph{IEEE Transactions on Pattern Analysis and
  Machine Intelligence}} \bibinfo{volume}{33}, \bibinfo{number}{1}
  (\bibinfo{year}{2011}), \bibinfo{pages}{117--128}.
\newblock


\bibitem[Joachims(2006)]%
        {joachims2006training}
\bibfield{author}{\bibinfo{person}{Thorsten Joachims}.}
  \bibinfo{year}{2006}\natexlab{}.
\newblock \showarticletitle{Training linear SVMs in linear time}. In
  \bibinfo{booktitle}{\emph{Proceedings of the 12th ACM SIGKDD International
  Conference on Knowledge Discovery and Data Mining}}.
  \bibinfo{pages}{217--226}.
\newblock


\bibitem[Kingma and Ba(2017)]%
        {kingma2017adam}
\bibfield{author}{\bibinfo{person}{Diederik~P. Kingma} {and}
  \bibinfo{person}{Jimmy Ba}.} \bibinfo{year}{2017}\natexlab{}.
\newblock \bibinfo{title}{Adam: A Method for Stochastic Optimization}.
\newblock
\newblock
\showeprint[arxiv]{1412.6980}~[cs.LG]


\bibitem[Malkov and Yashunin(2020)]%
        {hnsw2020}
\bibfield{author}{\bibinfo{person}{Yu~A. Malkov} {and} \bibinfo{person}{D.~A.
  Yashunin}.} \bibinfo{year}{2020}\natexlab{}.
\newblock \showarticletitle{Efficient and Robust Approximate Nearest Neighbor
  Search Using Hierarchical Navigable Small World Graphs}.
\newblock \bibinfo{journal}{\emph{IEEE Transactions on Pattern Analysis and
  Machine Intelligence}} \bibinfo{volume}{42}, \bibinfo{number}{4}
  (\bibinfo{date}{4} \bibinfo{year}{2020}), \bibinfo{pages}{824--836}.
\newblock


\bibitem[McNemar(1947)]%
        {mcnemartest1947}
\bibfield{author}{\bibinfo{person}{Quinn McNemar}.}
  \bibinfo{year}{1947}\natexlab{}.
\newblock \showarticletitle{Note on the sampling error of the difference
  between correlated proportions or percentages}.
\newblock \bibinfo{journal}{\emph{Psychometrika}}  \bibinfo{volume}{12}
  (\bibinfo{date}{6} \bibinfo{year}{1947}), \bibinfo{pages}{153--157}.
\newblock


\bibitem[Nguyen et~al\mbox{.}(2016)]%
        {nguyen2016msmarco}
\bibfield{author}{\bibinfo{person}{Tri Nguyen}, \bibinfo{person}{Mir
  Rosenberg}, \bibinfo{person}{Xia Song}, \bibinfo{person}{Jianfeng Gao},
  \bibinfo{person}{Saurabh Tiwary}, \bibinfo{person}{Rangan Majumder}, {and}
  \bibinfo{person}{Li Deng}.} \bibinfo{year}{2016}\natexlab{}.
\newblock \showarticletitle{MS MARCO: A Human Generated MAchine Reading
  COmprehension Dataset}.
\newblock  (\bibinfo{date}{November} \bibinfo{year}{2016}).
\newblock


\bibitem[Qin et~al\mbox{.}(2010)]%
        {qin2010general}
\bibfield{author}{\bibinfo{person}{Tao Qin}, \bibinfo{person}{Tie-Yan Liu},
  {and} \bibinfo{person}{Hang Li}.} \bibinfo{year}{2010}\natexlab{}.
\newblock \showarticletitle{A general approximation framework for direct
  optimization of information retrieval measures}.
\newblock \bibinfo{journal}{\emph{Information Retrieval}} \bibinfo{volume}{13},
  \bibinfo{number}{4} (\bibinfo{year}{2010}), \bibinfo{pages}{375--397}.
\newblock


\bibitem[Reimers and Gurevych(2019)]%
        {reimers-gurevych-2019-sentence}
\bibfield{author}{\bibinfo{person}{Nils Reimers} {and} \bibinfo{person}{Iryna
  Gurevych}.} \bibinfo{year}{2019}\natexlab{}.
\newblock \showarticletitle{Sentence-{BERT}: Sentence Embeddings using
  {S}iamese {BERT}-Networks}. In \bibinfo{booktitle}{\emph{Proceedings of the
  2019 Conference on Empirical Methods in Natural Language Processing and the
  9th International Joint Conference on Natural Language Processing
  (EMNLP-IJCNLP)}}. \bibinfo{pages}{3982--3992}.
\newblock


\bibitem[Robertson et~al\mbox{.}(1994)]%
        {bm25original}
\bibfield{author}{\bibinfo{person}{Stephen~E. Robertson},
  \bibinfo{person}{Steve Walker}, \bibinfo{person}{Susan Jones},
  \bibinfo{person}{Micheline Hancock-Beaulieu}, {and} \bibinfo{person}{Mike
  Gatford}.} \bibinfo{year}{1994}\natexlab{}.
\newblock \showarticletitle{Okapi at TREC-3.}. In
  \bibinfo{booktitle}{\emph{TREC}} \emph{(\bibinfo{series}{NIST Special
  Publication}, Vol.~\bibinfo{volume}{500-225})},
  \bibfield{editor}{\bibinfo{person}{Donna~K. Harman}} (Ed.).
  \bibinfo{publisher}{National Institute of Standards and Technology (NIST)},
  \bibinfo{pages}{109--126}.
\newblock


\bibitem[Rudin(2009)]%
        {Rudin:JMLR:2009}
\bibfield{author}{\bibinfo{person}{Cynthia Rudin}.}
  \bibinfo{year}{2009}\natexlab{}.
\newblock \showarticletitle{The P-Norm Push: A Simple Convex Ranking Algorithm
  That Concentrates at the Top of the List}.
\newblock \bibinfo{journal}{\emph{Journal of Machine Learning Research}}
  \bibinfo{volume}{10} (\bibinfo{date}{Dec.} \bibinfo{year}{2009}),
  \bibinfo{pages}{2233--2271}.
\newblock


\bibitem[Rudin and Wang(2018)]%
        {RuWa:aistats:2018}
\bibfield{author}{\bibinfo{person}{Cynthia Rudin} {and} \bibinfo{person}{Yining
  Wang}.} \bibinfo{year}{2018}\natexlab{}.
\newblock \showarticletitle{Direct Learning to Rank and Rerank}. In
  \bibinfo{booktitle}{\emph{Proceedings of Artificial Intelligence and
  Statistics {AISTATS}}}.
\newblock


\bibitem[Taylor et~al\mbox{.}(2008)]%
        {Taylor+al:2008}
\bibfield{author}{\bibinfo{person}{Michael Taylor}, \bibinfo{person}{John
  Guiver}, \bibinfo{person}{Stephen Robertson}, {and} \bibinfo{person}{Tom
  Minka}.} \bibinfo{year}{2008}\natexlab{}.
\newblock \showarticletitle{SoftRank: Optimizing Non-smooth Rank Metrics}. In
  \bibinfo{booktitle}{\emph{Proceedings of the 1st International Conference on
  Web Search and Data Mining}}. \bibinfo{pages}{77--86}.
\newblock


\bibitem[Thorne et~al\mbox{.}(2018)]%
        {thorne2018fever}
\bibfield{author}{\bibinfo{person}{James Thorne}, \bibinfo{person}{Andreas
  Vlachos}, \bibinfo{person}{Christos Christodoulopoulos}, {and}
  \bibinfo{person}{Arpit Mittal}.} \bibinfo{year}{2018}\natexlab{}.
\newblock \showarticletitle{{FEVER}: a Large-scale Dataset for Fact Extraction
  and {VER}ification}. In \bibinfo{booktitle}{\emph{Proceedings of the 2018
  Conference of the North {A}merican Chapter of the Association for
  Computational Linguistics: Human Language Technologies, Volume 1 (Long
  Papers)}}. \bibinfo{pages}{809--819}.
\newblock


\bibitem[Tonellotto et~al\mbox{.}(2018)]%
        {tonellotto2018survey}
\bibfield{author}{\bibinfo{person}{Nicola Tonellotto}, \bibinfo{person}{Craig
  Macdonald}, {and} \bibinfo{person}{Iadh Ounis}.}
  \bibinfo{year}{2018}\natexlab{}.
\newblock \showarticletitle{Efficient Query Processing for Scalable Web
  Search}.
\newblock \bibinfo{journal}{\emph{Foundations and Trends in Information
  Retrieval}} \bibinfo{volume}{12}, \bibinfo{number}{4--5} (\bibinfo{date}{Dec}
  \bibinfo{year}{2018}), \bibinfo{pages}{319--500}.
\newblock


\bibitem[Xia et~al\mbox{.}(2008)]%
        {xia2008listwise}
\bibfield{author}{\bibinfo{person}{Fen Xia}, \bibinfo{person}{Tie-Yan Liu},
  \bibinfo{person}{Jue Wang}, \bibinfo{person}{Wensheng Zhang}, {and}
  \bibinfo{person}{Hang Li}.} \bibinfo{year}{2008}\natexlab{}.
\newblock \showarticletitle{Listwise approach to learning to rank: theory and
  algorithm}. In \bibinfo{booktitle}{\emph{Proceedings of the 25th
  International Conference on Machine Learning}}. \bibinfo{pages}{1192--1199}.
\newblock


\bibitem[Xu and Li(2007)]%
        {Jun+Hang:2007}
\bibfield{author}{\bibinfo{person}{Jun Xu} {and} \bibinfo{person}{Hang Li}.}
  \bibinfo{year}{2007}\natexlab{}.
\newblock \showarticletitle{AdaRank: A Boosting Algorithm for Information
  Retrieval}. In \bibinfo{booktitle}{\emph{Proceedings of the 30th Annual
  International ACM SIGIR Conference on Research and Development in Information
  Retrieval}}. \bibinfo{pages}{391--398}.
\newblock


\bibitem[Yang et~al\mbox{.}(2018)]%
        {yang2018hotpotqa}
\bibfield{author}{\bibinfo{person}{Zhilin Yang}, \bibinfo{person}{Peng Qi},
  \bibinfo{person}{Saizheng Zhang}, \bibinfo{person}{Yoshua Bengio},
  \bibinfo{person}{William Cohen}, \bibinfo{person}{Ruslan Salakhutdinov},
  {and} \bibinfo{person}{Christopher~D. Manning}.}
  \bibinfo{year}{2018}\natexlab{}.
\newblock \showarticletitle{{H}otpot{QA}: A Dataset for Diverse, Explainable
  Multi-hop Question Answering}. In \bibinfo{booktitle}{\emph{Proceedings of
  the 2018 Conference on Empirical Methods in Natural Language Processing}}.
  \bibinfo{pages}{2369--2380}.
\newblock


\end{thebibliography}

\end{document}